\documentclass[12pt]{article}

\usepackage{amssymb,amsmath}	
\usepackage{graphicx}			
\usepackage{subfigure}			
\usepackage{color}				
\usepackage{authblk}	 		
\usepackage{hyperref}			
\usepackage{cite}

\usepackage[font=small]{caption}
\usepackage[left=2.0cm,right=2.0cm,top=2.5cm,bottom=3.0cm,a4paper,headheight=15pt]{geometry}
\graphicspath{{./figures/}}

\def\ba{\begin{array}}
\def\ea{\end{array}}
\def\be{\begin{eqnarray}}
\def\ee{\end{eqnarray}}
\def\nn{\nonumber\\}

\def\la{\label}
\def\eq#1{\eqref{#1}}

\def\d{\delta}
\def\D{\Delta}
\def\e{\epsilon}

\def\l{\lambda}
\def\L{\Lambda}

\def\th{\theta}

\def\r{\rho}

\def\ta{\tau}

\def\O{\Omega}

\def\fr{\frac}

\def\bra{\left\langle}
\def\ket{\right\rangle}
\def\lb{\left[}

\def\ls{\left(}
\def\lp{\left.}
\def\rp{\right.}
\def\rb{\right]}

\def\rs{\right)}

\def\Tr{{\rm Tr}\,}

\def\text#1{{\rm #1}}

\hypersetup{
  colorlinks   = true,	
  urlcolor     = blue,	
  linkcolor    = red,	
  citecolor    = green	
}

\title{\bf Quantum entanglement in inflationary cosmology}
\author[1]{Seoktae Koh\thanks{kundol.koh@jejunu.ac.kr}}
\author[2,3]{Jung Hun Lee\thanks{junghun.lee@gist.ac.kr}}
\author[2,3,4]{Chanyong Park\thanks{cyong21@gist.ac.kr}}
\author[3]{Daeho Ro\thanks{daeho.ro@apctp.org}}
\affil[1]{\small \it Jeju National University, Jeju 63243, Korea}
\affil[2]{\small \it Department of Physics and Photon Science, Gwangju Institute of Science and Technology,  Gwangju, 61005, Korea}
\affil[3]{\small \it Asia Pacific Center for Theoretical Physics, Pohang 37673, Korea}
\affil[4]{\small \it Department of Physics, Postech, Pohang 37673, Korea}
\date{\today}

\begin{document}
\maketitle
\thispagestyle{empty}

\vspace{1cm}
\begin{abstract}

We investigate the holographic quantum entanglement of a visible universe in an inflationary cosmology. To do so, we consider an AdS space with a dS boundary which represents an expanding space in time. In an inflationary cosmology, there exists a natural entangling surface called a cosmic event horizon which divides a universe into visible and invisible parts. In this model, cosmic event horizon monotonically decreases and approaches a constant value proportional to the inverse of Hubble constant. We show that the quantum entanglement between the visible and invisible universes divided by cosmic event horizon decreases monotonically in time. After an infinite time evolution, it finally approaches a constant value which is proportional to the inverse square of the Hubble constant for a four-dimensional dS space.

\end{abstract}


\newpage
\tableofcontents
\setcounter{page}{1}


\section{Introduction}

Considerable attention has been paid to the quantum entanglement entropy which becomes an important physical concept to figure out quantum features of a variety of physics areas. Although the entanglement entropy is well defined in a quantum field theory (QFT) \cite{Holzhey:1994we,Vidal:2002rm,Latorre:2003kg,Casini:2004bw}, it is not easy to calculate it for an interacting QFT. In this situation, holography recently conjectured in the string theory \cite{Maldacena:1997re,Gubser:1998bc,Witten:1998qj,Witten:1998zw} which allows us to evaluate such a nontrivial entanglement entropy nonperturbatively even for a strongly interacting theory \cite{Ryu:2006bv,Ryu:2006ef,Hubeny:2007xt,Solodukhin:2008dh,Nishioka:2009un,Casini:2009sr,Myers:2010tj,Takayanagi:2012kg, VanRaamsdonk:2010pw,Casini:2011kv}. We investigate the quantum entanglement of an expanding system and of a inflationary cosmology by applying holography.

In order to calculate the entanglement entropy, one first divides a system into two subsystems, $A$ and $B$, and then the reduced density matrix of $A$ is defined as the trace the total density matrix over the other subsystem $B$. In this case, two subsystems are divided by an entangling surface and an observer living in $A$ cannot receive any information from $B$. This situation is very similar to the black hole \cite{Bombelli:1986rw,Srednicki:1993im}. An observer living at the asymptotic boundary can never get any information from the inside of the black hole horizon. Because of such a similarity, there were many attempts to understand the Bekenstein-Hawking entropy in terms of the entanglement entropy \cite{Brustein:2005vx,Emparan:2006ni,Cadoni:2007vf,Solodukhin:2011gn,Casini:2012ei,Klebanov:2012yf,Nishioka:2014kpa,Myers:2012ed,Nozaki:2013wia,Nozaki:2014hna,Caputa:2014vaa,Park:2015hcz,Kim:2016jwu}. Furthermore, the similarity between the black hole horizon and the entangling surface has led to a new and fascinating holographic formula to calculate the entanglement entropy on the dual gravity side. Although the holographic method has not been proved yet, it was checked that the holographic formula perfectly reproduces the known results of a two-dimensional conformal field theory (CFT) \cite{Calabrese:2004eu,Calabrese:2005zw,Calabrese:2009qy,Lewkowycz:2013nqa,Kim:2016hig,Kim:2017lyx,Narayanan:2018ilr}.

In a cosmological model described by a dS space \cite{Maldacena:2012xp,Liu:2012eea}, there exists a specific surface called a cosmic event horizon which is similar to the black hole horizon. An observer living at the center of a dS space cannot see the outside of cosmic event horizon and cosmic event horizon radiates similar to the black hole horizon. From the quantum entanglement point of view, cosmic event horizon naturally divides a universe into two subsystems. One is a visible universe which we can see in future and the other is called a invisible universe. In this case, the invisible universe indicates that an observer living in a visible universe cannot see the outside of cosmic event horizon even after infinite time evolution. In general, cosmic event horizon remains as a constant in the late inflation era. The cosmic event horizon similar to the black horizon provides a natural entangling surface dividing a universe into two parts. Although the visible and invisible universes are casually disconnected from each other, quantum correlation between them can still exist. Therefore, it would be interesting to investigate the quantum entanglement between the visible and invisible universes, which may give us new information about the outside of our visible universe and the effect of the invisible universe on the cosmology of the visible universe.

In order to investigate the quantum entanglement between two subsystems in the expanding universe, we take into account an AdS space with a dS boundary space \cite{Bucher:1994gb,Sasaki:1994yt}. The minimal surface extended to such an AdS space corresponds to the entanglement entropy of an expanding space defined at the boundary of the AdS space \cite{Maldacena:2012xp}. Before studying the entanglement entropy of a visible universe, we first consider a system whose boundary expands in time unlike cosmic event horizon. In this case, the entanglement entropy in the early time era increases by the square of the cosmological time $\ta$, whereas it in the late time era grows up exponentially by $e^{(d-2) H \ta}$ for a $d$-dimensional QFT. If we take cosmic event horizon as an entangling surface, the entanglement entropy shows a totally different behavior. The cosmic event horizon at $\ta=0$ is located at the equator of a $(d-1)$-dimensional sphere and it monotonically decreases as the cosmological time goes on. In the late inflation era, cosmic event horizon approaches a constant value proportional to the inverse of Hubble constant. Similarly, the corresponding entanglement entropy also monotonically decreases and approaches a constant value at $\ta=\infty$.

The rest of this paper is organized as follows: In Sec. \ref{sec:2}, we briefly review an AdS space with a dS boundary. On this background, we study the entanglement entropy of an expanding system for $d=2,3,4$ cases in Sec. \ref{sec:3}. In Sec. \ref{sec:4}, we introduce a cosmic event horizon and divide a universe into visible and invisible universes. On this background, we study the quantum correlation between the visible and invisible universes in the inflationary cosmology. Finally, we finish this work with concluding remarks in Sec. \ref{sec:5}.


\section{AdS space with a dS boundary} \label{sec:2}

Consider a $(d+1)$-dimensional AdS space which can be embedded into a $(d+2)$-dimensional flat manifold with two time signatures. Denoting the $(d+2)$-dimensional flat metric as 
\be
ds^2 = - dY_{-1}^2 - dY_{0}^2 + \d_{ij} dY^i dY^j ,
\ee
where $i$ and $j$ run from $1$ to $d$, the Lorentz group of this $(d+2)$-dimensional flat space is given by $SO(2,d)$. In order to obtain a $(d+1)$-dimensional AdS metric, we impose the following constraint
\be
- R^2 = - Y_{-1}^2 - Y_{0}^2 + \d_{ij} Y^i Y^j .
\ee
Then, the hyper-surface satisfying this constraint represents a $(d+1)$-dimensional AdS space with
an AdS radius $R$. Since the imposed constraint is also invariant under the $SO(2,d)$ transformation, the resulting AdS geometry becomes a $(d+1)$-dimensional space invariant under the $SO(2,d)$ transformation which is nothing but the isometry group of the AdS space. There exist a variety of parametrizations satisfying the above constraint. In this work, we focus on the parametrization which allows a $d$-dimensional de Sitter (dS) space at the boundary. Now, let us parametrize the coordinates of the ambient space as \cite{Maldacena:2012xp}
\be
Y_{-1} = R \cosh \fr{\r}{R} \quad , \quad
Y_{0} = R \sinh \fr{\r}{R} \sinh \fr{t}{R} \quad {\rm and} \quad Y^i = R  n^i \sinh \fr{\r}{R} \cosh \fr{t}{R}  ,
\ee
where $n^i$ indicates a $d$-dimensional orthonormal vector satisfying $\d_{ij} n^i n^j=1$. The resulting AdS metric then gives rise to
\be  		\la{res:dp1metric}
ds^2 = d\r^2 + \sinh^2 \ls \fr{\r}{R} \rs \lb - d t^2 + R^2 \cosh^2  \ls \fr{t}{R} \rs    \ls
d \th^2 + \sin^2 \th d \O_{d-2}^2 \rs \rb ,
\ee
where $d \th^2 + \sin^2 \th d \O_{d-2}^2$ indicates a metric of a $(d-1)$-dimensional unit sphere. According to the AdS/CFT correspondence, the boundary of this AdS space defined at $\r=\infty$ can be regarded as the space-time we live in. Above the boundary metric shows a dS space which can describes an inflationary cosmology. In this work, after dividing the boundary space into two subsystems, we investigate the quantum correlation between them.

In order to divide the boundary space into two subsystems, let us first assume that we are at $\th=0$, and that the two subsystem is bordered at $\th_o$. For convenience, we call the subsystem we are in is an observable system and the other subsystem an unobservable system. In general, the border is called the entangling surface in the entanglement entropy study. Although we do not get any information from the unobservable system, the quantum state of the observable system can be affected from the unobservable system due to the nontrivial quantum entanglement. Assuming that the entire system is given by a pure state $\lp |\Psi \ket$ represented by the product of two subsystem's states \cite{Ryu:2006bv,Ryu:2006ef,Casini:2011kv,Calabrese:2004eu,Calabrese:2005zw,Calabrese:2009qy,Rosenhaus:2014woa,Rosenhaus:2014ula,Rosenhaus:2014zza}
\be
\lp |\Psi \ket = \lp |\psi \ket_o \lp |\psi \ket_{u} ,
\ee
where $\lp |\psi \ket_o$ and $\lp |\psi \ket_{u}$ indicate the state of the observable and unobservable systems, respectively. Then, the reduced density matrix of the observable system is given by tracing over the unobservable part
\be
\r_o = \Tr_{u}  \lp |\Psi \ket \bra  \Psi | \rp ,
\ee
and the entanglement entropy is described by the Von Neumann entropy
\be
S_E = - \Tr_o \ \r_o \log \r_o .
\ee
Although the entanglement entropy is conceptually well defined in a quantum field theory, it is not easy to calculate it in general cases. Recently, Ryu and Takayanagi have proposed a new method called the holographic entanglement entropy \cite{Ryu:2006bv,Ryu:2006ef}. According to the AdS/CFT correspondence, the entanglement entropy can be easily evaluated by calculating the minimal surface area extended to the dual geometry. Following the holographic proposition, we will discuss the entanglement entropy of the expanding system in \eq{res:dp1metric}.

\section{Entanglement entropy on the expanding system } \label{sec:3}

Until now, we have discussed about the entanglement entropy between the observable and unobservable systems. However, it is not still clear how we can divide the observable and unobservable systems. One simple choice is to take a constant $\th_o$. Under this simple ansatz, the sizes of the two subsystems gradually increases as time goes on. The set-up with a constant $\th_o$ may be useful to describe an expanding material or to figure out the entanglement entropy of a time-dependent subsystem. On the other hand, it is also interesting to take into account the time-dependent $\th_o $. In this section, we first investigate the entanglement entropy defined by a constant $\th_o$ and then discuss further the entanglement entropy in the inflationary cosmology  with a time-dependent $\th_o$ in the next section

For simplicity, let us first consider the $d=2$ case which can give us a solvable toy model. For $d=2$, the dual geometry reduces to the three-dimensional AdS space
\be		\la{res:d2metric}
ds^2 = d\r^2 + \sinh^2 \ls \fr{\r}{R} \rs  \lb - d t^2 +  R^2 \cosh^2 \ls \fr{t}{R} \rs   \ d \th^2 \rb .
\ee
If we focus on the boundary space of this AdS space with a fixed $\r$, the boundary metric has the form of the cosmological type metric depending on time. In order to describe the entanglement entropy on the time-dependent background, we assume that the observable system is in the range of
\be
- \fr{\th_o}{2} \le \th \le \fr{\th_o}{2} .
\ee
In this section, we regard $\th_o$ as a constant as mentioned before. Then, $\pm \th_o/2$ correspond to two boundaries of the observable system. Since we took the constant $\th_o$, the size of the observable system expands. More precisely, the size of the observable system is given by $\sinh (\L/R)  \cosh (t/R) R \th_o$ at the AdS boundary denoted by $\r=\L$. This shows that the size of the observable system increases by $\cosh (t/R)$. Since $\cosh (t/R)$ is invariant under the time reversal, from now on we take into account only the non-negative time period, $0 \le t < \infty$. This implies that the observable system begins the expansion at $t=0$. If we take $\L$ as an infinity, it usually leads to a divergence. In the holographic set-up, this divergence is associated with a UV divergence of the dual field theory and $\L$ is introduced to regularize the UV divergence. If we take a finite but large energy scale for $\L$, it may be associated with the energy scale at which the expansion begins.

In order to get more physical intuition about the entanglement entropy on the time-dependent geometry, let us consider several particular limits. We first define the turning point as $\r_*$ which corresponds to the minimum value extended by the minimal surface. In the case with $ \r_* /R \gg 1$, we can calculate the entanglement entropy analytically but perturbatively even for higher dimensional cases. This parameter range corresponds to the UV limit and may give rise to a good guide line to figure out physical implication for the numerical study. For $\r_* /R \gg 1$, the entanglement entropy is governed by \cite{Park:2015afa,Park:2015dia,Kim:2014yca,Kim:2014qpa,Kim:2018mgz}
\be
S_E = \fr{1}{4 G} \int_{-\th_o/2}^{\th_o/2} d \th \sqrt{\r'^2 + \fr{ R^2}{4}   e^{2 \r/R} \cosh^2 \ls t /R \rs  } .
\ee
Solving the equation of motion derived from it, $\th_o$ at given $t$ is determined by the turning point
\be
\th_o = \fr{4 }{e^{\r_*/R} \cosh ( t/R )} .
\ee
When $\th_o$ and $t$ are given, inversely, the turning point can also be regarded as a function of $\th_o$ and $t$
\be		\la{res:sturnpt}
e^{\r_*/R} = \fr{4}{\th_o } \fr{1}{ \cosh ( t/R ) } .
\ee
Note that $t/R$ must not be large to obtain a large $\r_*/R$. This fact implies that the approximation with $\r_*/R \gg 1$ is valid only in the early time. In addition, this result shows that the turning point goes into the interior of the AdS space as time evolves.

Performing the integral of the entanglement entropy with the obtained solution, the resulting entanglement entropy reduces to
\be
S_E = \fr{ \lb  (\L- \r_*) + R \log 2 \rb }{2 G } .
\ee
This result together with \eq{res:sturnpt} shows that the entanglement entropy increases by $t^2$ for $t/R \ll 1$
\be		\la{res:earlybe1}
S_E \sim  \frac{  R \log \th_o + \Lambda - R\log 2  }{2 G}+\frac{ t^2  }{4 G R} .
\ee
If $ t/R > 1$, on the other hand, it increases linearly in time
\be   		\la{res:earlybe2}
S_E \sim \frac{ R  \log \th_o + \Lambda -2 R \log 2  }{2 G }+\frac{ t}{2 G}  .
\ee

Now, let us take into account a more general case without the constraint $ \r_*/R \gg 1$. The general form of the entanglement entropy reads from \eq{res:d2metric}
\be
S_E = \fr{1}{4 G} \int_{-\th_o/2}^{\th_o/2} d \th \sqrt{\r'^2 + R^2 \sinh^2 ( \r/R) \cosh^2 (t/R)  } .
\ee
After solving the equation of motion, performing the integral gives rise to
\be
\fr{\th_o}{2} &=& \int_{\r_*}^{\infty} d \r \ \frac{ \sinh \left(\rho _*  /R \right)  }{ R \cosh(t/R) \sinh ( \rho/R ) \sqrt{\sinh ^2(\rho/R )-\sinh ^2\left(\rho _*/R \right)}} \nn
&=& \fr{1}{ \cosh  (t/R)  } \lb \fr{\pi}{2}  - \arctan \ls \sinh ( \r_*/R)  \rs \rb .
\ee
Rewriting it leads to the following relation
\be
\sinh (\r_*/R) = \cot \ls \fr{\th_o \cosh (t/R)}{2}\rs ,
\ee
which reproduces the previous result in \eq{res:sturnpt} for $\r_* /R \gg 1$. In general case, the resulting entanglement entropy reads
\be		\la{res:exact2dHEE}
S_E 
&=& \frac{\Lambda }{2 G}
+\frac{  R \log \lb \sin \left(\frac{1}{2} \th_o \cosh (t/R) \right)\rb}{2 G }.
\ee
When $\th_o \cosh (t/R) \ll 1$, this result again reproduces the previous ones obtained in the early inflation era.

\begin{figure}
\begin{center}
\includegraphics[width=0.45\textwidth]{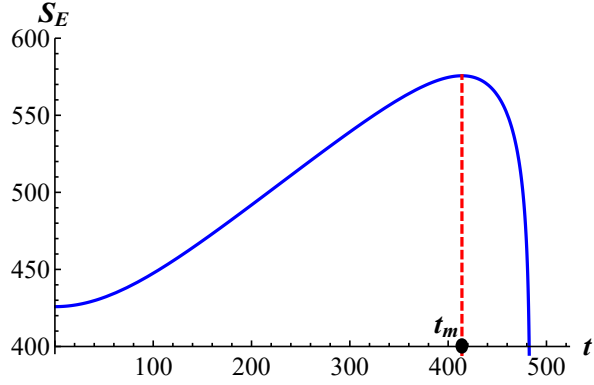}
\caption{We take $\epsilon=1/1000$, $R=100$, $\th_o = 1/10$ and $G=1$.}
\label{d2afig1}
\end{center}
\end{figure}

It is worth noting that the resulting entanglement entropy is well defined only in the time range of $0 \le t < t_f$, where $t_f$ satisfies $\th_o \cosh (t_f /R)= 2 \pi$. After this critical time $t_f$, the logarithmic term of the entanglement entropy is not well defined. Now, let us define an additional critical time $t_m$ satisfying $\th_o \cosh (t_m/R) =  \pi$. At this critical time ($t=t_m$), the observable and unobservable systems have the same size. In this case, the turning point is located at $\r_*=0$ and the entanglement entropy has a maximum value, $S_E=\L/(2 G)$. Near $t_m$ ($t<t_m$), the entanglement entropy approaches this maximum value slowly  by $- (t_m - t)^2$
\be
S_E \approx  \frac{  \Lambda }{2 G} -\frac{\th_o^2  \sinh ^2\left( t_m /R \right)}{16 G R} \   (t_m - t)^2   + {\cal O} \ls (t_m - t)^4 \rs .
\ee
From these results, we can see that the entanglement entropy of the observable system increases by $t^2$ in the early time and saturates the maximum value at a finite time $t_m$. After $t_m$, the entanglement entropy rapidly decreases as shown in Fig. \ref{d2afig1}. As a consequence, we can summarize the entanglement entropy of an one-dimensional expanding system as follows

\begin{itemize}
\item In the early time with $\r_*/R \gg1$ and $t/R \ll1$, the entanglement entropy increases by $t^2$.
\item In the intermediate era with $\r_*/R \gg1$ and $t/R > 1$, the entanglement entropy increases linearly as time evolves.
\item In the late time with $ \r_*/R \sim 0$ and $t \approx t_m$, the entanglement entropy slowly increases by $- (t_m - t)^2$ and finally saturates the maximum value at $t=t_m$.
\item After $t_m$, the entanglement entropy rapidly decreases.
\end{itemize}
In Fig. \ref{d2afig1}, we plot the exact entanglement entropy given in \eq{res:exact2dHEE}, which shows the time dependence expected by the analytic calculation in several particular limits.

It has been well known that the entanglement entropy of a two-dimensional CFT dual to an AdS$_3$ has a logarithmic divergence and its coefficient is proportional to the central charge of the dual CFT \cite{Ryu:2006bv,Ryu:2006ef}. However, the above result for AdS$_3$ with the dS$_2$ boundary shows a linear divergence ($\sim \L$) instead of the logarithmic one. This is because the coordinate used in this work is different from the one usually used in Ref. \cite{Ryu:2006bv,Ryu:2006ef,Narayanan:2018ilr}. To see this, let us introduce a new coordinate
\be
\sinh ( \r/R) =  \fr{R}{ z}  .
\ee
Then, the three-dimensional AdS metric in \eq{res:d2metric} can be rewritten as
\be
ds^2 = \fr{R^2 d z^2}{ z^2 (1+z^2/R^2)}  + \fr{R^2}{ z^2} \lb - d t^2 + R^2 \cosh^2 ( t/R) \  d \th^2  \rb ,
\ee
where the boundary is located at $z=0$. In the UV limit ($z \to 0$) with $t/R \ll1$, the new coordinate is related to the original one by $e^{ \r/R} \sim R/z$ and the above metric is further simplified to
\be
ds^2 \approx \fr{R^2 d z^2}{ z^2 }  + \fr{R^2}{z^2} \lb - d t^2 + \fr{R^2}{4} \ d \th^2  \rb ,
\ee
which is locally equivalent to the AdS space in the Poincare patch. Thus, the linear divergence appearing in \eq{res:exact2dHEE} can be reinterpreted as a logarithmic one in the new coordinate system, $\L/R = - \log  (\e/R)$, where $\e$ indicates the UV cut-off of the $z$-coordinate. As a result, the linear divergence obtained here is consistent with the known logarithmic one up to the coordinate transformation.

\subsection{On higher dimensional expanding observable system}

Now, let us take into account higher dimensional cases with $d \ge 3$. For convenience, we use the new coordinate $R/ z =\sinh ( \r/R)$. Then, the previous $(d+1)$-dimensional AdS metric can be rewritten as
\be		\la{metric:general}
ds^2 = \fr{R^2 d z^2}{z^2 (1+z^2/R^2)}  + \fr{R^2}{z^2} \lb - d t^2 + R^2 \cosh^2 (t/R)  \  \ls
d \th^2 + \sin^2 \th d \O_{d-2}^2 \rs \rb ,
\ee
where the boundary is located at $z=0$. On this background, the holographic entanglement entropy is governed by
\begin{equation}			\la{eq:entangleform}
S_E = \dfrac{\Omega_{d-2} R^{2 d-3} \cosh^{d-2} (t/R) }{4  G} \int_{0}^{\th_o} d\theta  \ \frac{\sin^{d-2}\theta }{z^{d-1}} \sqrt{\dfrac{z'^2}{1 + z^2/R^2} + R^2 \cosh^2 (t/R) } ,
\end{equation}
where we take the range of $\th$ as $0 \le \th \le \th_o$ instead of $-\th_o /2 \le \th \le \th_o/2$. Varying this action at a given time, the configuration of a minimal surface is determined by the highly nontrivial differential equation. Since it does not allow us to write an analytic solution, the numerical study is inevitable for a higher dimensional theory. However, if we focus on the early time behavior of the entanglement entropy, we can find a perturbative and analytic solution and gain more intuitions. In this section, we first study analytically the entanglement entropy in the early time and then look into the time evolution of the entanglement entropy numerically.

Now, let us discuss the entanglement entropy of the observable system in the early time with $t/R \ll 1$. We first assume that the observable system is very tiny in the early time. Then, we can take $\th_o \ll 1$. In this case, the minimal surface is extended only to the UV region represented as $0 \le z \le z_*$ with $z_*/R \ll 1$. This is because $z_*/R$ is usually proportional to $\th_o$ at $t=0$, as will be seen. Due to the small size of the observable system, the AdS metric in the early time can be well approximated by
\be			\la{metric:pert}
ds^2 \approx \fr{R^2 dz^2}{z^2 (1+z^2/R^2)} + \fr{R^2}{z^2} \lb - d t^2 + R^2 \cosh^2 ( t/R) \    \ls  d \th^2 + \th^2  d \O_{d-2}^2 \rs \rb .
\ee
On this background, the entanglement entropy is given by
\be		\la{act:original}
S_E = \frac{\Omega_{d-2}  R^{2d-3} \cosh ^{d-2} (t/R)}{4   G} \int_{0}^{\th_o} d \th \ \fr{\theta ^{d-2}  }{z^{d-1}  }  \sqrt{ \fr{z'^2}{1+z^2/R^2} + R^2 \cosh ^2 (t/R) }.
\ee

In order to find a perturbative solution satisfying $z/R \le z_*/R \ll 1$, we introduce a small parameter $\l$ for indicating the smallness of the solution. Then, the perturbative expansion of the solution can be parametrized as
\be		\la{ansatz:pertsol}
z(\theta)= \l \ls z_0 (\theta) + \l z_1(\theta) + \l^2 z_2 (\theta)+\cdots \rs .
\ee
When varying this perturbative solution with respect to $\th$, it is worth noting that the derivative of the solution, $z'(\theta)$, must be expanded as
\be
z'(\theta)= z_0' (\theta) + \l z_1'(\theta) + \l^2 z_2'(\theta)+\cdots  .
\ee
This is because $\th$ has the same order of $z/R$ in the early time. Before performing the explicit calculation, let us think about the parity transformation, $z \to - z$ and $\th \to - \th$. Under this parity transformation, we can easily see that the metric in \eq{metric:pert} and  the entanglement entropy are invariant. If we transforms $\l \to - \l$ instead of $z_n$ in \eq{ansatz:pertsol}, only $z_{2n}$ terms give rise to the consistent transformation with $z \to - z$. Due to this reason, the $z_{2n+1}$ terms automatically vanish. As a consequence, we can set $z_1(\th)=0$ without loss of generality.

At leading order of $\l$, the entanglement entropy is given by
\be			\la{act:generalact}
S_0 =  \frac{\Omega_{d-2}  R^{2 d-3} \cosh ^{d-2} (t/R)}{4   G} \int_{0}^{\th_o} d \th \ \fr{\theta ^{d-2}  }{z_0^{d-1}  }   \sqrt{ z_0'^2 + R^2  \cosh ^2 (t/R)  } .
\ee
In a higher dimensional theory unlike the $d=2$ case, the entanglement entropy relies on $\th$ explicitly. Thus, there is no well-defined conserved quantity unlike the $d=2$ case. This fact implies that we must solve the second order differential equation to obtain the entanglement entropy. At leading order, the minimal surface configuration can be determined by solving the equation of motion derived from $S_0$
\be			\la{eq:generaleq}
0 &=& \fr{2  \cosh ^2(t/R) \theta  z_0  z_0'' }{R^2} + \fr{2 (d-2) z_0  z_0'^3}{R^4 } + \fr{2  (d-1)  \cosh ^2(t/R)  \theta  z_0'^2}{R^2 }  \nn
&&  + \fr{2 (d-2)   \cosh ^2(t/R)  z_0  z_0' }{R^2} +   2 (d-1)  \cosh ^4 ( t/R)  \theta  .
\ee
Despite the complexity of the equation of motion, it allows the following simple and exact solution regardless of the dimension $d$
\be			\la{res:leadingsol}
\fr{z_0}{R} =  \cosh (t/R)  \sqrt{\th_o^2 - \th^2} .
\ee
From this, we see that the turning point denoted by $z_*$ is proportional to $\th_o$, as mentioned before,
\be
\fr{z_*}{R} = \th_o  \cosh (t/R)  .
\ee
Note that this relation is derived from the leading order entanglement entropy. If we further consider higher order corrections, the turning point can vary with some small corrections.

When a UV cut-off denoted by $\e$ is given, we can easily see from the background metric that the volume of the observable system is given by
\be
{\cal V}_{d-1} =  \fr{\O_{d-2} R^{2(d-1)} \cosh^{d-1} (t/R)}{d-1} \ \fr{\th_o^{d-1}}{\e^{d-1}} ,
\ee
while the area of the entangling surface becomes
\be
{\cal A}_{d-2} (t)  =  \O_{d-2} R^{2(d-2)} \cosh^{d-2} (t/R) \ \fr{\th_o^{d-2}}{\e^{d-2}} .
\ee
These formulae show that the area of the entangling surface increases by $\cosh^{d-2} (t/R)$ as time evolves. At $t=0$, in particular, the area reduces to
\be			\la{res:coshoarea}
\bar{{\cal A}}_{d-2} =   \O_{d-2} R^{2(d-2)}  \ \fr{\th_o^{d-2}}{\e^{d-2}} ,
\ee
which can be determined by two parameters, $\e$ and $\th_o$. In the holographic study, the minimal surface is extended only to $\e \le z \le z_*$, so that  $z_*> \e $ must be satisfied for consistency. Recalling further that $ z_*/R = \th_o$ at $t=0$, we finally obtain $\th_o > \e/R$. This fact implies that, when the expansion begins at $t=0$, the observable system and the entangling surface have the non-vanishing volume and area.

Now, let us consider the $d=3$ case. Using the perturbative expansion discussed before, the entanglement entropy is expanded into
\be
S_E = S_0 + S_2 + \cdots,
\ee
with
\be
S_0 &=& \fr{\O_1 R^3 \cosh (t/R)}{4  G} \int_0^{\th_o-\th_c}  d \th \ \fr{\th}{ z_0^2}  \sqrt{  z_0'^2 + R^2 \cosh^2 (t/R)  } , \nn
S_2 &=& - \fr{\Omega _1 R^4 \cosh (t/R) }{8 G } \int_0^{\th_o}  d \th \ \frac{\theta  \lb z_0^3 z_0'^2 + 4 R^2  z_2 z_0'^2- 2 R^2 z_0 z_0' z_2' +4 R^4 z_2 \cosh ^2 (t/R) \rb }{ R^3 z_0^3 \sqrt{z_0'^2+R^2 \cosh ^2 (t/R)}} ,
\ee
where we set $\l=1$ and introduce $\th_c$ as a UV cut-off in the $\th$-direction. In the second integral, $\th_c$ was removed because it does not give any additional UV divergence. Substituting the leading order solution in  \eq{res:leadingsol} into $S_0$ and performing the integral, we finally obtain the leading contribution to the entanglement entropy
\be
S_0 = \frac{\Omega _1 R^2  \sqrt{\th_o} }{4 \sqrt{2} G  \sqrt{\theta_c}}-\frac{\Omega _1 R^2}{4  G } .
\ee

The first correction caused by $z_2(\th)$ is determined by the following differential equation
\be
0  = z_2''+ \frac{\left(\th_o^2 - 2 \theta ^2\right) }{\theta  \left(\th_o^2 - \theta ^2 \right)} z_2'   -\frac{2 \th_o^2 }{\left(\th_o^2 - \theta ^2 \right){}^2} z_2 + 2  R \sqrt{\th_o^2-\theta ^2} \cosh ^3(t/R)  .
\ee
This equation allows an exact solution
\be
z_2 =c_2 -\frac{c_2 \th_o \tanh ^{-1}\left(\frac{\sqrt{\th_o^2-\theta ^2}}{\th_o}\right)}{\sqrt{\th_o^2-\theta
   ^2}}+\frac{6 c_1 +\left(\theta ^4-4 \th_o^2 \theta ^2+3 \th_o^4+4 \th_o^4 \log \theta  \right) R \cosh^3(t/R)}{6  \sqrt{\th_o^2-\theta ^2} } ,
\ee
where $c_1$ and $c_2$ are two integral constants. These two integral constants must be fixed by imposing two appropriate boundary conditions. The natural boundary conditions are $z_2 (\th_o) = 0$ and $z'(0)=0$. The first conditions implies that the end of the minimal surface is located at the boundary, while the second constraint is required to obtain a smooth minimal surface at $\th=0$. These two boundary conditions determine two integral constants to be
\be
c_1 &=&  - \fr{2 \th_o^4 R \log\th_o \cosh^3 (t/R) }{3 } ,\nn
c_2 &=&   - \fr{2 \th_o^3 R \cosh^3 (t/R)}{3 } .
\ee
Substituting the found perturbative solutions again into $S_2$, the first correction to the entanglement entropy is given by
\be
S_2 = - \fr{5 \th_o^2 \O_1 R^2 \cosh^2 (t/R)}{36  G } .
\ee

Above the regulator $\th_c$ is usually associated with the regulator $\e$ in the $z$-direction. Using the perturbative solution we found, $\th_c$ can be represented as a function of $\e$
\be
\th_c =  \fr{\e^2 }{ 2 \th_o R^2  \cosh^2 (t/R)}  - \fr{2    \e^3}{9 R^3 \cosh (t/R) } +  {\cal O} (\e^4)
\ee
As a consequence, the resulting perturbative entanglement entropy leads to
\be
S_E =\frac{\th_o \Omega_1 R^3 \cosh (t/R)}{4 G  \e}-\frac{\Omega_1   R^2  \left(\th_o^2  \cosh ^2 (t/R)+3 \right)}{12 G } + {\cal O} \ls \e \rs.
\ee
Recalling the formula in \eq{res:coshoarea}, this entanglement entropy can be rewritten as
\be
S_E =\frac{ {{\cal A}}_1 (t)  R}{4 G}-\frac{\Omega_1  R^2  \left(\th_o^2 \cosh ^2 (t/R)+3 \right)}{12 G } + {\cal O} \ls \e \rs ,
\ee
where ${\cal A}_1 (t)$ indicates the area of the entangling surface at a given time $t$. The leading contribution to the entanglement entropy, as expected, satisfies the area law even in the time-dependent space. Expanding it further in the early time, the entanglement entropy leads to
\be
S_E =  \frac{ \bar{\cal A}_1 R}{4 G } -\frac{\Omega _1 R^2}{4 G }
-\frac{\th_o^2 \Omega _1R^2}{12 G } + \ls \frac{\bar{\cal A}_1}{8 G R}
-\frac{\th_o^2  \Omega _1}{12 G} \rs t^2
 + {\cal O} \ls t^4 \rs ,
\ee
where $\bar{\cal A}_1 = {\cal A}_1 (0)$. This result shows that the entanglement entropy in the early time increases by $t^2$
\be
S_E (t) - S_E(0) \approx \ls \frac{\bar{\cal A}_1}{8 G R} -\frac{\th_o^2  \Omega _1}{12 G} \rs t^2  .
\ee
It also shows that the increase of the entanglement entropy is proportional to the area of the entangling surface at leading order.

\begin{figure}
\begin{center}
\hspace{-0.0cm}
\subfigure[][]{\includegraphics[width=0.45\textwidth]{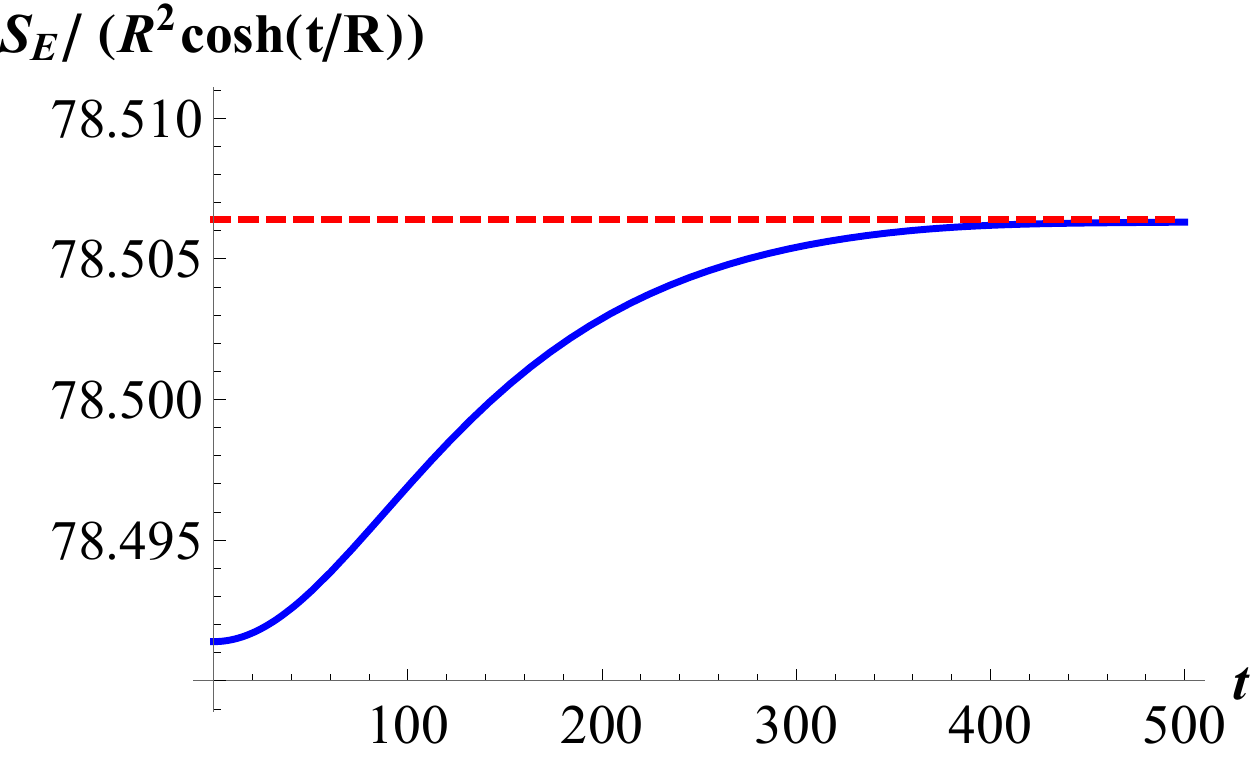}}
\hspace{-0.0cm}
\subfigure[][]{\includegraphics[width=0.45\textwidth]{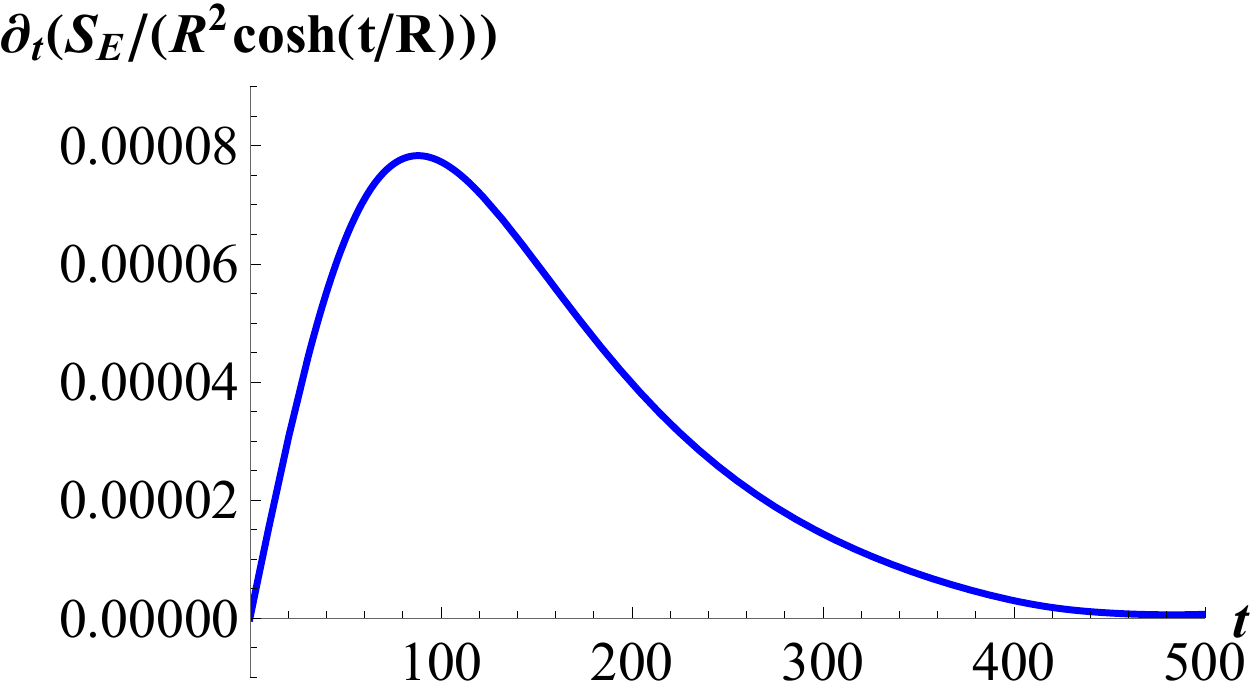}}
\caption{We take $\epsilon=1/1000$, $R=100$, $\th_o = 1/20$ and $G=1$.}
\label{d3nfig1}
\end{center}
\end{figure}

\begin{figure}
\begin{center}
\subfigure{\includegraphics[width=0.45\textwidth]{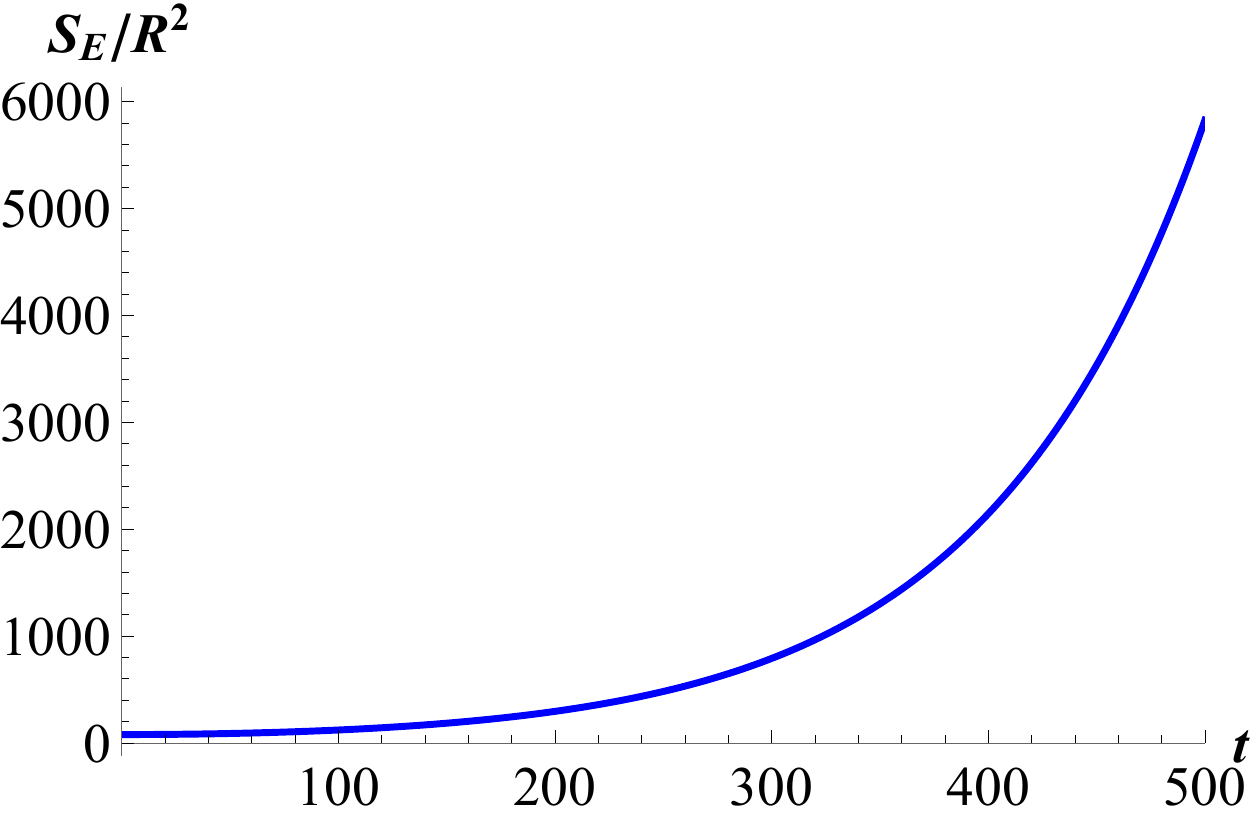}}
\caption{We take $\epsilon=1/1000$, $R=100$, $\th_o = 1/20$ and $G=1$ .}
\label{d3nfig0}
\end{center}
\end{figure}

In order to see the entanglement entropy in the late time, we must go beyond the perturbative expansion. After finding a numerical solution satisfying \eq{eq:generaleq}, we investigate how the corresponding entanglement entropy increases in time. In Fig. \ref{d3nfig1}, we depict the value of $S_E / \ls R^2  \cosh (t/R) \rs$ and its time derivative. In Fig. \ref{d3nfig1}(a), the value of $S_E  /\ls R^2 \cosh (t/R) \rs$ approaches a constant in the late time. This fact becomes manifest in Fig. \ref{d3nfig1}(b), where the time derivative of $S_E / \ls R^2  \cosh (t/R) \rs$ approaches zero in the late time. Consequently, we can see that the entanglement entropy increases exponentially ($S_E \sim e^{t/R}$) in the late time (see Fig. \ref{d3nfig0}).

\begin{figure}
\begin{center}
\hspace{-0.0cm}
\subfigure[][]{\includegraphics[width=0.45\textwidth]{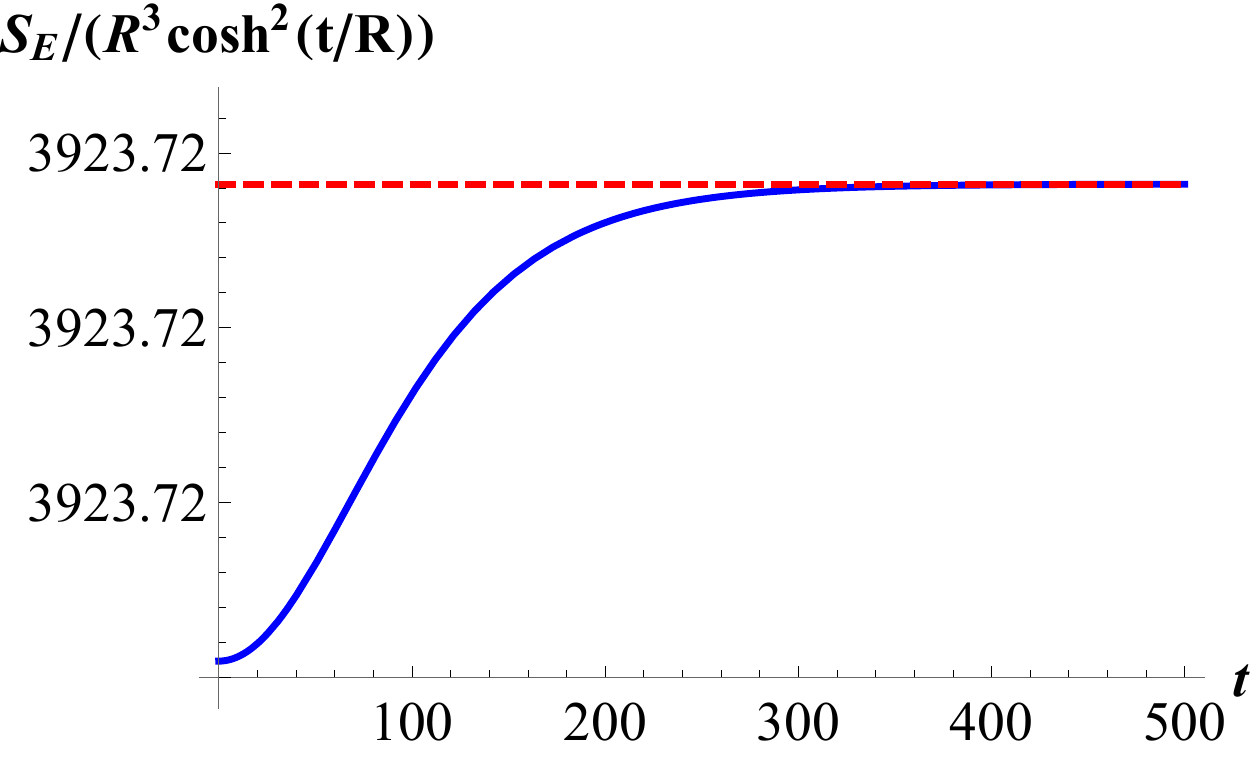}}
\hspace{-0.0cm}
\subfigure[][]{\includegraphics[width=0.45\textwidth]{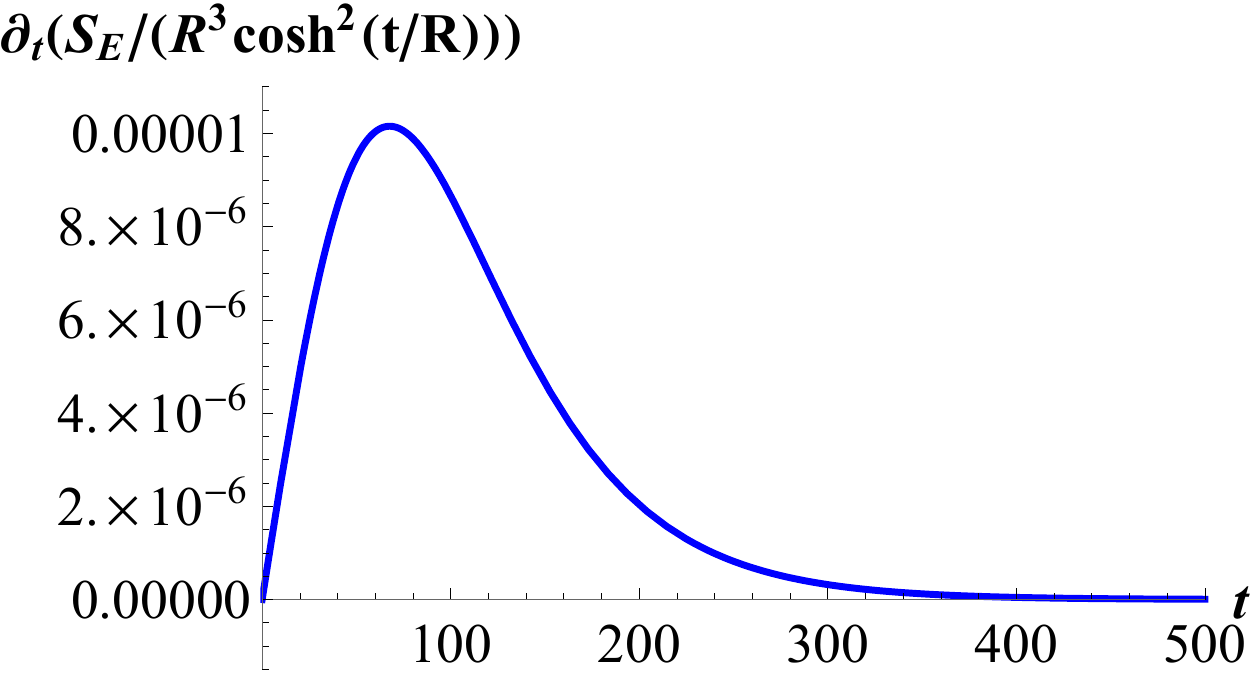}}
\caption{We take $\epsilon=1/1000$, $R=100$, $\th_o = 1/20$ and $G=1$.}
\label{d4nfig1}
\end{center}
\end{figure}

Repeating the same calculation for $d=4$, the entanglement entropy of the $d=4$ observable system, similar to the $d=3$ case, increases by $t^2$ in the early time and exponentially grows in the late time. In the late time, the increment of the entanglement entropy is proportional to $S_E \sim e^{t/R}$ for $d=3$ and $S_E \sim e^{2t/R}$ for $d=4$ which becomes manifest in Fig. \ref{d4nfig1}. These results imply that the entanglement entropy of the expanding observable system increases by $t^2$ in the early time regardless of $d$ and in the late time grows by $S_E \sim e^{(d-2) t/R}$ for a general $d$. For the black hole formation corresponding to the thermalization of the dual field theory, the entanglement entropy usually increases by $t^2$ in the early time similar to the expanding observable system. However, in the late time of the thermalization the entanglement entropy is saturated and becomes a thermal entropy, while the entanglement entropy of the expanding observable system increases exponentially in the late time.

\section{Entanglement entropy of the visible universe in the inflationary cosmology} \label{sec:4}

In the previous section, we studied the quantum entanglement of the expanding observable system which is described by the constant $\th_o$. In this section, we investigate the entanglement entropy of the visible universe in the inflationary cosmology. In an inflationary model, there exists a natural way to divide the entire universe into two parts. Because of the growing scale factor in the inflationary model, there exists an invisible universe which we cannot see forever. On the other hand, the universe we can see is called the visible universe and the boundary of the visible universes is called cosmic event horizon which corresponds to the border of the visible and invisible universes. In this case, the invisible universe is casually disconnected from us. Due to the existence of the natural border of two universes in the inflationary model, it would be interesting to study the quantum correlation between them. In this section, we will investigate such an entanglement entropy for a four-dimensional inflationary cosmology.

\begin{figure}
\begin{center}
\subfigure[][]{\includegraphics[width=0.45\textwidth]{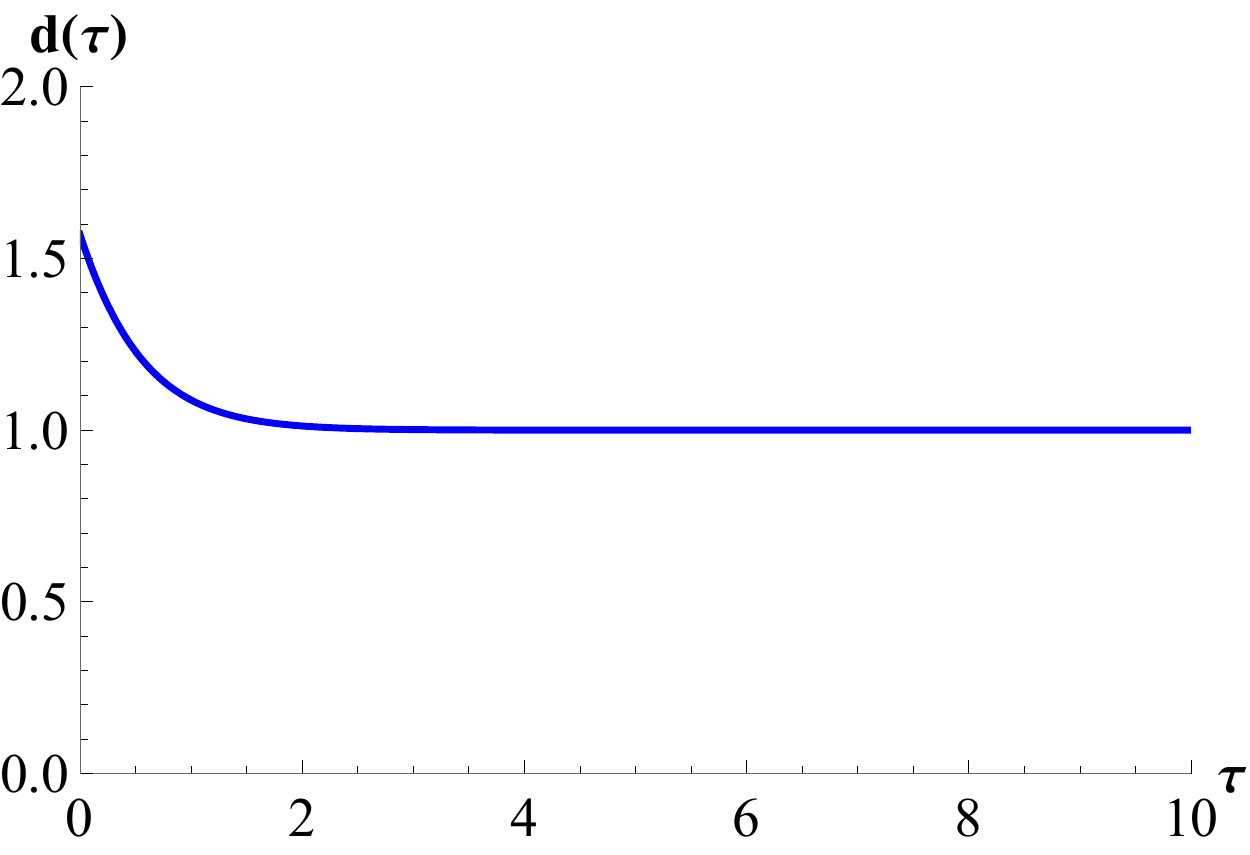}}
\hspace{0.5cm}
\subfigure[][]{\includegraphics[width=0.45\textwidth]{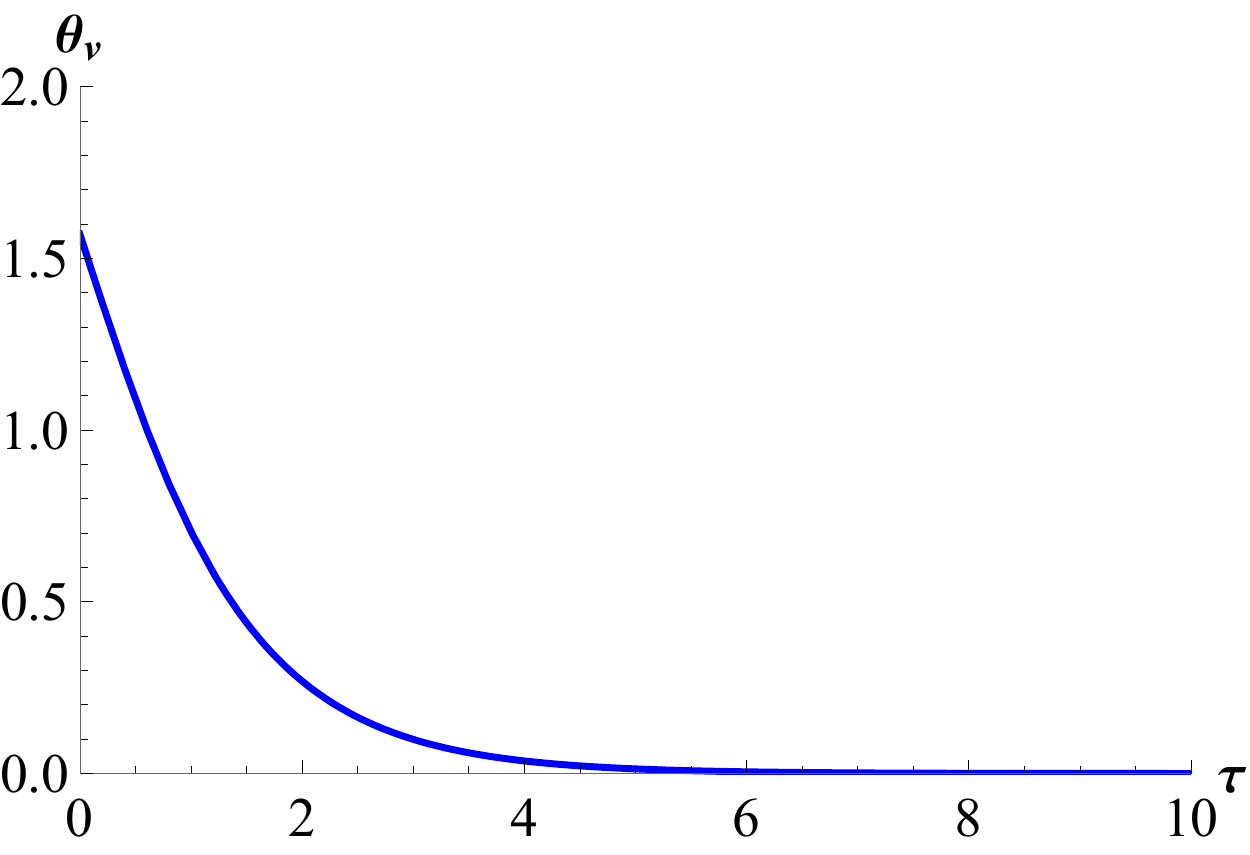}}
\caption{cosmic event horizon relying on the cosmological time $\ta$ where we take $H=1$.}
\label{d4nfig2}
\end{center}
\end{figure}

Let us first define cosmic event horizon as the boundary of the visible universe. From \eq{metric:general} for $d=4$, the boundary metric reads at $z=\e$
\be
ds_B^2 = \fr{R^2}{ \e^2} \lb - d t^2 + R^2 \cosh^2 (t/R)   \  \ls
d \th^2 + \sin^2 \th d \O_{d-2}^2 \rs \rb ,
\ee
which describes ${\bf R}^+ \times {\bf S}^3$.
In order to interpret the boundary metric as the cosmological one, we introduce a cosmological time $\ta$ and Hubble constant $H$ such that
\be			\la{rel:timeH}
\ta = \fr{R}{\e} t \quad {\rm and} \quad H = \fr{\e}{R^2} .
\ee
Then, the boundary metric reduces to the one representing an inflationary cosmology
\be
ds_B^2 =   - d \ta^2 + \fr{\cosh^2 (H \ta)}{H^2}   \  \ls d \th^2 + \sin^2 \th d \O_{d-2}^2 \rs   ,
\ee
where the scale factor is given by $a(\ta)= \cosh  ( H  \ta)  /H  $. Due to the nontrivial scale factor, the distance travelled by light is restricted to a finite region whose boundary by definition corresponds to cosmic event horizon. More precisely, cosmic event horizon in the above cosmological metric is determined by
\be			\la{res:ceventh}
d(\ta) = a(\ta) \int_t^{\infty}  \fr{c \ d \ta'}{a(\ta')} = \lb \fr{\pi}{2}  - 2 \arctan \ls \tanh \fr{H \ta}{2} \rs  \rb \fr{\cosh H \ta}{H},
\ee
where the light speed was taken to be $c=1$. In Fig. \ref{d4nfig2}(a), we plot how cosmic event horizon changes as the cosmological time $\ta$ evolves. In the early inflation era, cosmic event horizon decreases as time goes on, whereas it approaches a constant value $1/H$ in the late inflation era which is a typical feature of the dS space.

The existence of cosmic event horizon indicates that the visible universe, the inside of cosmic event horizon, is casually disconnected from the invisible universe, the outside of cosmic event horizon \cite{Gibbons:1977mu,MargalefBentabol:2013bh,Anderson:1983nq,Anderson:1984jf}. In other words, if we are at the center of the visible universe, we can never receive any information from the invisible universe. Even in this situation, there can exist a nontrivial quantum correlation between them, which can be measured by the entanglement entropy. From the viewpoint of the entanglement entropy, cosmic event horizon naturally plays a role of an entangling surface which divides a system into two subsystems. Therefore, it would be interesting to investigate the entanglement entropy of the inflationary cosmology to know how the our visible universe is quantumly correlated to the invisible universe we cannot see forever.

To go further, let us re-express cosmic event horizon in terms of the angle appearing in the AdS space. For distinguishing cosmic event horizon from the previous expanding entangling surface parametrized by $\th_o$, we use a different symbol $\th_v$ which is given by a function of $\ta$ unlike $\th_o$. Assuming that we are at the north pole of the three-dimensional sphere denoted by $\th=0$, our visible universe can be characterized by $0 \le \th \le \th_v$. In this case, the radius of the entangling surface is determined from the AdS metric
\be
l = \int_0^{\th_v} d \th \ \fr{\cosh   H  \ta  }{H} = \fr{\th_v \cosh   H  \ta  }{H}  .
\ee
Because the radius of the entangling surface must be identified with cosmic event horizon, the comparison of them determines $\th_v$ as a function of the cosmological time
 \be			\la{rel:cosmicthv}
\tan \ls \fr{\pi}{4} - \fr{\th_v}{2} \rs  = \tanh \fr{H \ta}{2} .
\ee
This result shows that $\th_v$ start with $\pi/2$ at $\ta=0$ and gradually decreases to $0$ at $\ta=\infty$ with a fixed subsystem size $l$. In the late inflation era, cosmic event horizon becomes a constant independent of the cosmological time, $d(\ta) = 1/H$. In Fig. \ref{d4nfig2}(b), we plot $\th_v$ relying on the cosmological time. In this figure, $\th_v$ starts from $\pi/2$ at $\ta=0$ and monotonically and rapidly decreases to $0$ as the cosmological time goes on.

By using $\th_v$ we found, it is possible to calculate holographically the entanglement entropy of the visible universe. Before performing the calculation, it is worth noting that the cosmological time and the Hubble constant are defined only at the boundary. The minimal surface corresponding to the entanglement entropy of the visible universe is extended to the bulk of the dual geometry, so that we cannot exploit the definition of $\ta$ and $H$ in the course of calculating the area of the minimal surface. After the calculation, however, we can replace $t$ and $\e$ with $\ta$ and $H$ through \eq{rel:timeH}. This is because the resulting area of the minimal surface represents the entanglement entropy defined at the boundary at which $\ta$ and $H$ are well defined.

\subsection{Entanglement entropy at \texorpdfstring{$\ta=0$}{tau=0}}

For simplicity, let us first consider the entanglement entropy at $\ta=0$. Using the relation in \eq{rel:timeH}, $\ta=0$ implies $t=0$ regardless of $\e$. For $d=4$, the holographic entanglement entropy formula is given by \eq{eq:entangleform} with $t=0$ and $\th_v$ instead of $\th_o$. If we alternatively take into account $\th$ as a function of $z$, the corresponding entanglement entropy in the inflationary model can be rewritten as
\be
S_E = \dfrac{\Omega_{2} R^{5} }{4  G} \int_{\e}^{\infty} d z  \ \frac{\sin^{2}\theta }{z^{3}} \sqrt{ R^2 \dot{\th}^2  + \dfrac{1}{1 + z^2/R^2}  } ,
\ee
where the dot indicates a derivative with respect to $z$. Deriving the equation of motion from this action, it allows a specific solution which satisfies $\dot{\th}=0$ and furthermore $\th = \pm \pi/2$. This solution indicates an equatorial plane of $S^{3}$. Performing the above integral with this equatorial plane solution, we finally obtain
\be
S_E = \dfrac{\Omega_{2} R^{5} }{4  G}  \ls \frac{1}{2 \epsilon ^2}  -\frac{1}{2 R^2}  \log  \fr{2 R}{\e} + \fr{1}{4 R^2}  \rs .
\ee
If we interpret $\e$ as the UV cut-off, this result shows the power-law divergence together with the logarithmic divergence, as expected in the entanglement entropy calculation for $d=4$. Rewriting $\e$ in terms of $H$ by using \eq{rel:timeH}, we finally obtain the following entanglement entropy at $\ta=0$
\be			\la{res:HEEatt0}
S_E = \dfrac{\Omega_{2} R }{8  G H^2}   - \dfrac{\Omega_{2} R^{3} }{8  G}  \log  \fr{2}{H R} + \dfrac{\Omega_{2} R^{3} }{16  G}    .
\ee

\subsection{Entanglement entropy in the late inflation era}

In the inflationary cosmology unlike the previous expanding system, the perturbative calculation of the entanglement entropy is possible in the late inflation era because $\th_v$ becomes small at large $t$ or $\ta$. In the late inflation era we can apply the previous perturbative expansion of $z$. Using the perturbation of $z$, the leading contribution and the first correction to the entanglement entropy are given by
\be
S_0 &=& \fr{\O_2 R^5 \cosh^2 (t/R)}{4  G} \int_0^{\th_v-\th_c}  d \th \ \fr{\th^2}{ z_0^3}  \sqrt{  z_0'^2 +R^2  \cosh^2 (t/R)  } , \\
S_2 &=& - \fr{\Omega _2 R^6 \cosh^2 (t/R) }{8 G } \int_0^{\th_v-\th_c}  d \th \ \frac{\theta  \left(z_0^3 z_0'^2+6 R^2 z_2 z_0'^2-2 R^2 z_0 z_0' z_2'+6 R^4 z_2 \cosh ^2 (t/R)  \right)}{ z_0^4 R^3 \sqrt{z_0'^2+R^2 \cosh ^2 (t/R) }} . \nonumber
\ee
Note that unlike the $d=3$ case, the upper limit of the integral range in $S_2$ has $\th_c$.
This is because we need to reintroduce $\th_c$ to regularize an additional divergence appearing in $S_2$ for $d=4$. Substituting the leading solution in \eq{res:leadingsol} into $S_0$, we obtain the following leading contribution to the entanglement entropy
\be
S_0 = \fr{\th_v R^3 \O_2}{16 G  \th_c} - \fr{\O_2 R^3}{16 G } \log \fr{2 \th_v}{\th_c} - \fr{\O_2 R^3 }{32 G } .
\ee
In this result, we can see that, when $\th_c \to 0$, the leading contribution leads to the expected power-law and logarithmic divergences for $d=4$.

Now, let us consider the deformation of the minimal surface described by $z_2$, which is governed by the following differential equation
\be
0= z_2'' +\frac{2 }{\theta } z_2'-\frac{3 \th_v^2  }{\left(\th_v^2-\theta ^2\right){}^2} z_2
+\frac{\left(3 \th_v^2-2 \theta ^2\right) R \cosh ^3  (t/R) }{\sqrt{\th_v^2-\theta ^2} } .
\ee
This equation allows us to find the following exact solution
\be
z_2 &=& \frac{c_1 \left(\th_v -\theta\right) ^2 }{\theta  \sqrt{\th_v^2-\theta ^2}}+\frac{c_2}{\sqrt{\th_v^2-\theta ^2}} +\fr{(\theta ^5-5 \th_v^2 \theta ^3-2 \th_v^3  \theta ^2-2 \th_v^4 \theta -2 \th_v^5) R \cosh ^3  (t/R) }{6 \theta \sqrt{\th_v^2-\theta ^2}  } \nn
&& +\frac{\left\{ \left(\th_v +\theta \right){}^2 \log \left(\th_v +\theta \right) -   \left(\th_v -\theta\right){}^2  \log \left(\theta
   _0-\theta \right) \right\}  \th_v^3 R \cosh ^3  (t/R) }{2 \theta \sqrt{\th_v^2-\theta ^2} } ,
\ee
where $c_1$ and $c_2$ are two integration constants. Imposing two boundary condition, $z_2 (\th_v) = 0$ and $z'(0)=0$ discussed in the previous section, $c_1$ and $c_2$ are determined to be
\be
c_1 &=& \fr{\th_v^3 R \cosh^3  (t/R) }{3 } , \nn
c_2 &=&  \fr{\th_v^4 R \cosh^3  (t/R)  \lb5 - 6 \log (2 \th_v) \rb}{3} .
\ee
Substituting the obtained solutions into $S_2$ again and performing the integral result in
\be
S_2 = -\frac{3 \th_v^2\Omega_2 R^3 \cosh^2  (t/R) }{32G}
\log\frac{2\th_v}{\theta_c}+ \frac{11 \th_v^2\Omega_2 R^3 \cosh^2  (t/R) }{64G}  .
\ee
When $\th_c \to 0$, it shows that the first correction gives rise to an additional logarithmic divergence unlike the known entanglement entropy.

\begin{figure}
\begin{center}
\subfigure{\includegraphics[width=0.45\textwidth]{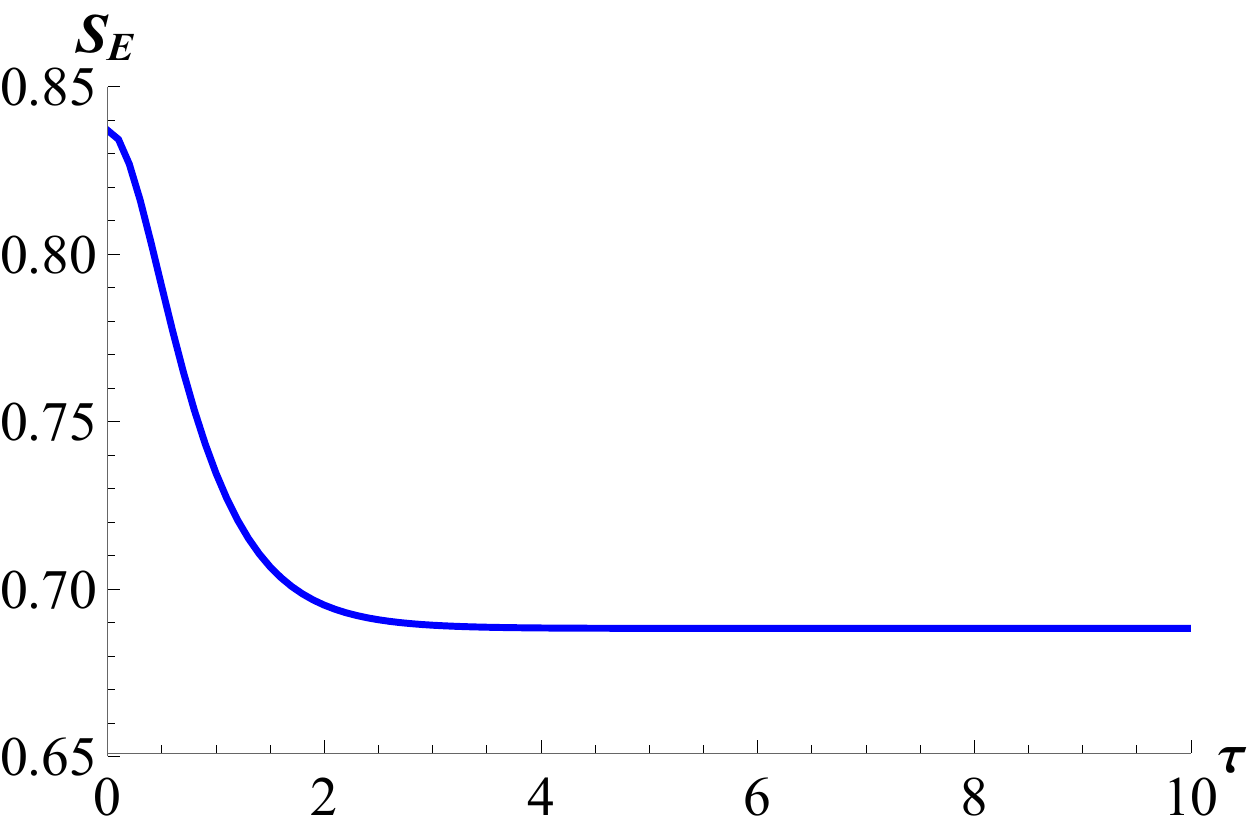}}
\caption{We take $\epsilon=1$, $R=1$ and $G=1$ for simplicity.}
\label{d4ent}
\end{center}
\end{figure}

From the solutions obtained perturbatively, $\th_c$ is determined in terms of $\e$
\be
\th_c &=&  \fr{\e^2}{ 2 \th_v R^2 \cosh^2  (t/R) }
+\frac{\e^4 }{8 \th_v^3 R^4 \cosh^4  (t/R) }  + \fr{\e^4}{48 \th_v R^4 \cosh^2  (t/R) }   \nn
&&  - \fr{\e^4}{4 \th_v  R^4 \cosh^2  (t/R) }  \log \fr{2 \th_v R \cosh  (t/R) }{ \e} + {\cal O} \ls \e^6\rs .
\ee
Using this relation, the resulting entanglement entropy leads to
\be
S_E &=&\frac{R {\cal A}_2 (t) }{8 G  }
-\frac{\Omega_2 R^3 }{16 G} \log  \fr{4 {\cal A}_2 (t)  }{\O_2 R^2}  -\frac{\Omega _2 R^3}{16 G } \nn
   && +   \fr{\th_v^2 \Omega _2 R^3 \cosh ^2 (t/R) }{6 G }   - \fr{\th_v^2 R^3 \Omega _2 \cosh ^2  (t/R) }{16 G}  \log \fr{4 {\cal A}_2 (t)  }{\O_2 R^2}  ,
\ee
where the area of cosmic event horizon is given by
\be
{\cal A}_2 (t) = \fr{\th_v^2 R^4 \Omega _2 \cosh ^2  (t/R) }{ \e^2} .
\ee
Replacing $t$ and $\e$ by $\ta$ and $H$ by using \eq{rel:timeH}, $\th_v$ and the area of cosmic event horizon in the late inflation era ($H\ta \gg 1$) are approximated by
\be
\th_v &\approx& 2 e^{-H \ta} , \nn
{\cal A}_2 (\ta) &\approx& \fr{\O_2}{H^2} .
\ee
As a result, the entanglement entropy of the visible universe in the late inflation era leads to the following expression
\be			\la{res:HEEinLIE}
S_E &\approx&\frac{\O_2 R  }{8 G  H^2}
-\frac{\Omega_2 R^3 }{4 G} \log  \fr{2 }{ R H}   +   \fr{ 5 \Omega _2 R^3 }{48 G }    .
\ee
This result shows that the entanglement entropy of the visible universe in the late inflation era is time-independent and determined by the Hubble constant and the area of cosmic event horizon. This is because cosmic event horizon remains as a constant in the late inflation era. The change of the entanglement entropy during the inflation era is given by
\be
\D S_E \equiv S_E (\infty) - S_E(0)  =
-\frac{\Omega_2 R^3 }{8 G} \log  \fr{2 }{ R H}   +   \fr{  \Omega _2 R^3 }{24 G }   ,
\ee
where the result in \eq{res:HEEatt0} was used. Since $HR = \e/R \ll 1$, $\D S_E$ always becomes negative. This indicates that the quantum correlation between the visible and invisible universes decreases with time. In Fig. \ref{d4ent}, we plot how the entanglement entropy of the visible universe changes as the cosmological time goes on. As expected by the perturbative and analytic calculation, the entanglement entropy gradually decreases and finally approaches a constant value after an infinite time.

\section{Discussion} \label{sec:5}

In this work, we have studied the quantum entanglement entropy of the expanding system and the inflationary universe. In order to take into account the expanding system and universe holographically, we considered an AdS space whose boundary is given by a dS space. In order to describe the quantum entanglement on the expanding system and space, we have investigated the holographic entanglement entropy of a subsystem on the boundary of the AdS space.

In this model, we took two different subsystems. One of them corresponds to an expanding system in which we determined the subsystem size with a fixed $\th_o$. In this case, since the volume of the boundary space increases with the cosmological time, the subsystem size also increases. In the early time era, we found that the entanglement entropy of an expanding system increases by $t^2$ regardless of the dimensionality of the system. In the late time era, on the other hand, we showed that, when the boundary of the AdS space expands with the expansion rate of $e^{H t}$, the increase of the entanglement entropy of a $d$-dimensional system is proportional to $e^{(d-2) t}$ for a $d$-dimensional space-time.

For a dS space, there is an important length scale called cosmic event horizon. If an observer is at the center of a dS space, he cannot see the outside of cosmic event horizon even after the infinite time evolution. In other words, the observer at the center of dS can never get any information from the outside of cosmic event horizon. From the quantum information viewpoint, cosmic event horizon like a black hole horizon resembles the entangling surface dividing a total system into two subsystems. In the present model, cosmic event horizon starts with $\th_v=\pi/2$ at $\ta=0$ and eventually approaches $\th_v=0$ at $\ta=\infty$ with a fixed $\th_v e^{H \ta}$. In the late inflation era, cosmic event horizon is given by the inverse of the Hubble constant, $d (\infty)= 1/H$. We showed that the entanglement entropy of the visible universe in the inflationary cosmology decreases continuously as time evolves and that it finally approaches a finite value independent of the cosmological time.


\vspace{1cm}

\section*{\small Acknowledgement}
S. Koh (NRF-2016R1D1A1B04932574), J. H. Lee (NRF-2016R1A6A3A01010320), C. Park (NRF-2016R1D1A1B03932371) and D. Ro (NRF-2017R1D1A1B03029430) were supported by Basic Science Research Program through the National Research Foundation of Korea funded by the Ministry of Education. J. H. Lee, C. Park and D. Ro were also supported by the Korea Ministry of Education, Science and Technology, Gyeongsangbuk-Do and Pohang City.


\bibliographystyle{ref_jhep}
\bibliography{ref}

\end{document}